# We need to talk about nonprobability samples

Robin J. Boyd[1], Gary D. Powney[1], Oliver L. Pescott[1]

[1]UK Centre for Ecology and Hydrology, Benson Lane, Crowmarsh Gifford, Oxfordshire, UK, OX10 8BB

Corresponding author email: robboy@ceh.ac.uk

Author twitter handles: @roboyd91, @GaryPowney and @sacrevert

## Abstract

In most circumstances, probability sampling is the only way to ensure unbiased inference about population quantities where a complete census is not possible. As we enter the era of "big data", however, *non*probability samples, whose sampling mechanisms are unknown, are undergoing a renaissance. We explain why the use of nonprobability samples can lead to spurious conclusions, and why seemingly large nonprobability samples can be (effectively) very small. We also review some recent controversies surrounding the use of nonprobability samples in biodiversity monitoring. These points notwithstanding, we argue that nonprobability samples can be useful, provided that their limitations are assessed, mitigated where possible and clearly communicated. Ecologists can learn much from other disciplines on each of these fronts.

Key words: biodiversity monitoring; convenience sample; risk-of-bias; sample representativeness; citizen science; selection bias

## Highlights

- As the data revolution gathers pace, researchers are increasingly relying on nonprobability samples from meta-databases, citizen science and other sources to monitor biodiversity.
- The use of nonprobability samples can lead to biased inference, and seemingly large nonprobability samples can actually have very low information content.
- A number of recent high-profile disagreements in the biodiversity literature stem from the use of such samples, and the inadequate communication of their potential weaknesses.
- Nonprobability samples can be useful for the purpose of monitoring biodiversity, provided that their limitations are assessed, mitigated where possible, and the almost inevitable remaining issues clearly communicated.

## Monitoring the biodiversity crisis

There is a scientific consensus that the sixth mass extinction of life on earth is underway [1]. To understand the scale of the problem, data on the state of biodiversity, and how it has changed over time, are needed. Collecting and analysing such data is known as biodiversity monitoring, and is an active area of research. As we will argue, however, biodiversity monitoring often rests on shaky statistical foundations.

## There is no census of life on earth

Monitoring biodiversity is typically a matter of **descriptive statistical inference** (see Glossary). There is an implied finite population, which comprises all observation units of interest. These units might be, say, patches of land across a landscape. The researcher wants to infer something about those population units; for example, the average species richness. Putting measurement error to one side, it is simple to calculate this quantity if each patch of land has been sampled. In many cases, however, it is not possible to census all population units. In such situations, researchers rely on a sample, and use sample-based **estimators** of the population quantities of interest. In the above example, it would be typical to use the sample mean as an estimator of the population mean.

## Probability samples, nonprobability samples and estimator bias

Broadly speaking, statisticians define two types of sample: probability samples and nonprobability samples. In a probability sample, the probability that each population unit was included in the sample is known. The simplest type of probability sampling is Simple Random Sampling (SRS), in which each population unit has an equal chance of selection [2]. In a nonprobability sample, the sampling mechanism, and therefore the chance that each population unit was sampled, is not known *a priori*.

Many sample-based estimators of population quantities—including sample means and proportions— are known to be **unbiased** under SRS [2]. A key property of SRS is that, as sample size increases, they are likely to be **representative** of the population (note the distinction between the variable representativeness of a single SRS, and the long-run unbiasedness of such samples over many hypothetical realisations). By representative, we mean that there is little to no correlation between an indicator variable, taking the value 1 if the population unit is in the sample and 0 otherwise, and the variable of interest [3,4]. If this correlation is ~0, the values of the variable of interest in the sample are similar to those in the population, and, consequently, sample averages and proportions etc. are similar to their population equivalents.

In addition to SRS, it is relatively straightforward to construct unbiased estimators for other types of probability sample [2]. Recall that in a probability sample, the probability that each population unit was sampled is known by design. Again, where sample size is not small, these probabilities can be used to correct for inbuilt unrepresentativeness. For example, rather than estimating a population total using a sample total, one would instead use the weighted total, where the weights are equal to the inverse of the sample selection probabilities [2]. Weights of this type are known as **design weights**, and using them to construct unbiased estimators is known as **design-based inference**.

Matters are more complicated for nonprobability samples. Nonprobability samples are often unrepresentative of the population, but estimators of population quantities are unlikely to be fully adjustable using inclusion probabilities because these are not precisely known. Methods have been developed to mitigate unrepresentativeness in nonprobability samples, which we review below (see "How can we do better?"), but it is typically very difficult to know how well this has been achieved [5].

## Quantity does not necessarily imply quality

Whilst it is challenging to construct unbiased estimators for nonprobability samples, they are frequently used for inferential research on pragmatic grounds [6]. Probability samples, even relatively small ones, are often very difficult to collect. On the other hand, nonprobability samples, even large ones, are relatively easy to come by. Hence, researchers often justify the use of nonprobability samples on the basis of data quantity.

Unfortunately, quantity of data is no substitute for representativeness. Meng [4] derived a formula for the "effective" size of a sample (Box 1). The effective size of a sample is equivalent to the size of the SRS that would yield an estimate of the population average with the same Mean Squared Error (MSE), assuming that the sample average is used as the estimator. An important implication of the formula is that, where a sample is even slightly unrepresentative, i.e. there is a correlation between sample membership and the variable of interest, the effective sample size becomes much smaller than actual sample size (Box 1). The formula also shows that as the sampling rate (sample size divided by population size) increases, bias (i.e. non-sampling error, a.k.a. systematic or irreducible error) comes to dominate random sampling error.

**Box 1.** Selection bias and data "bigness".

Meng [4] derived a formula relating selection bias to the accuracy of the sample average $\bar{Y}_n$ as an estimator of the population average $\bar{Y}_N$. It is not possible to provide the full derivation here, we simply present two relevant equations. The first shows that the difference between the sample average and the population average is:

$$(\bar{Y}_n) - (\bar{Y}_N) = \rho(R,Y) \sqrt{\frac{1-f}{f}} \sigma_Y$$

where $\rho(R,Y)$ is the (population) correlation between Y and an indicator variable taking the value 1 if the population unit is in the sample and 0 otherwise, f is the sampling rate (n/N), and the final term $\sigma_Y$ is the population standard deviation of Y. $\rho(R,Y)$ indicates the direction and magnitude of the selection bias: when $\rho(R,Y) > 0$ larger values of Y are more likely in the sample than in the population, and vice versa. Where $\rho(R,Y) = 0$, this term cancels the others and there is no error.

The second equation gives the "effective" size $n_{eff}$ of a sample, defined as the size of SRS that would produce an estimate of the population average with the same mean squared error (on average). $n_{eff}$ may be expressed as:

$$n_{eff} = \frac{f}{(1-f)} \frac{1}{E[\rho(R,Y)^2]} = \frac{n}{1-f} \frac{1}{E[\rho(R,Y)^2]} \frac{1}{N}$$

$E[\rho(R,Y)^2]$ is the expectation of the square of $\rho(R,Y)$ for a given data selection mechanism. $E[\rho(R,Y)^2]$ is expressed as an expectation because $\rho(R,Y)$ has many possible realisations for a given mechanism. However, where N is large, the variance in $\rho(R,Y)$ is typically negligible so $E[\rho(R,Y)^2]$ can be substituted by $\rho(R,Y)^2$ [4].

An important implication of this formula is that, where $\rho(R,Y)^2$ deviates even slightly from 0, $n_{eff}/n$ (i.e. the relative effective sample size) decreases with N. In biodiversity monitoring, N is typically very large, so this reduction can be substantial. Thus, for nonprobability samples where $\rho(R,Y) \neq 0$, big data can be (effectively) very small. Figure I demonstrates the impact of selection bias on the effective sample size of a nonprobability biodiversity dataset.

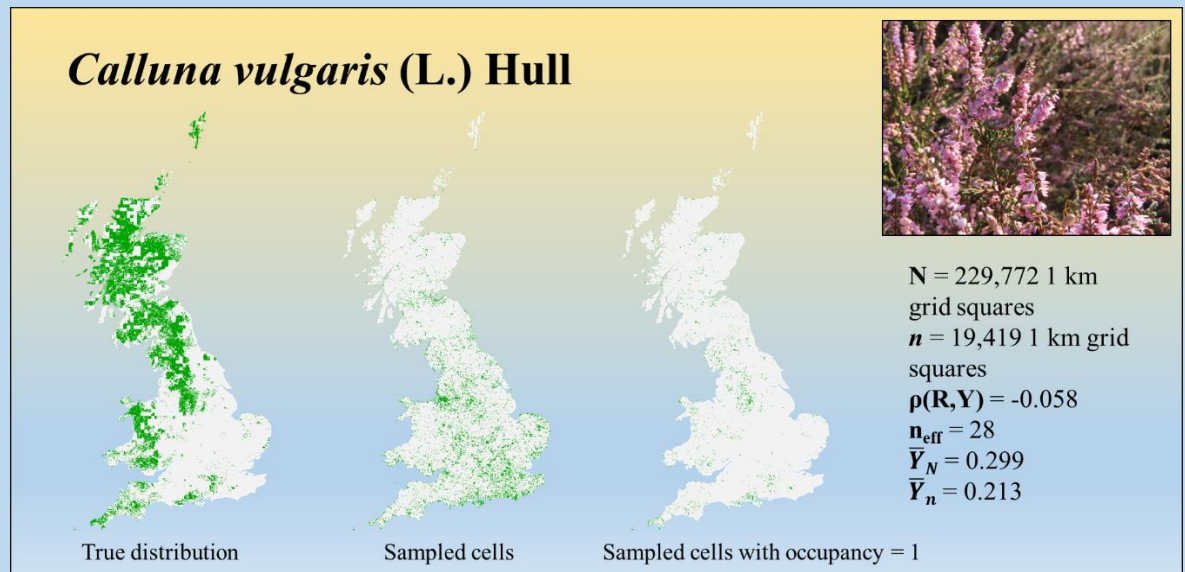

*Calluna vulgaris* (L.) Hull

N = 229,772 1 km grid squares
*n* = 19,419 1 km grid squares
ρ(R,Y) = -0.058
$n_{eff}$ = 28
$\bar{Y}_N$ = 0.299
$\bar{Y}_n$ = 0.213

True distribution  Sampled cells  Sampled cells with occupancy = 1

**Figure I.** The effects of selection bias on effective sample size. Here, the population comprises 229,772 1 km land-containing grid cells in Britain. 19,419 cells were sampled by volunteers, who submitted records of vascular plants to iRecord (www.irecord.org.uk) via the smartphone app, the website, or indirectly via the iSpot initiative (https://www.ispotnature.org/) from 2000–2019. The The formula for calculating effective sample size yields startling results. In his original paper, Meng [4] analysed a dataset of size n = 2,300,000 on voting intentions for the 2016 US election. He found that the correlation between sample membership and respondents' intention to vote for Trump was a modest -0.005. This seemingly miniscule correlation led to an effective sample size of ~400 (a 99.98% reduction). More recently, Bradley and colleagues [10] used the same formula to analyse several surveys of COVID-19 vaccine uptake in the US. They showed that the largest (but non-random) survey, with 250,000 responses per week, can produce estimates with the same error as a random sample of less than 10. Summarising, the authors stated, "*[o]ur central message is that data quality matters more than data quantity, and that compensating the former with the latter is a mathematically provable losing proposition.*"

Figure I (Box 1) illustrates the effect of selection bias on our ability to accurately estimate mean occupancy of the common heather *Calluna vulgaris* in Britain. The selection bias, i.e. the correlation between sample membership and occupancy, is -0.058. The sample-based estimate of mean occupancy is 0.213, a substantial underestimate of the population mean, 0.299. The effective sample size, 28, is 99.86% smaller than the actual sample size of 19,419!

The estimate of mean occupancy is not just wrong, but precisely wrong. The apparent sample size is large, which means that the normal approximation of the 95% confidence interval for the estimate of mean occupancy is narrow. Consequently, as we show in Box 2, it has virtually no chance of covering the true value. This is the big data paradox: "the more the data, the surer we are to fool ourselves." [4]

**Box 2.** The big data paradox.

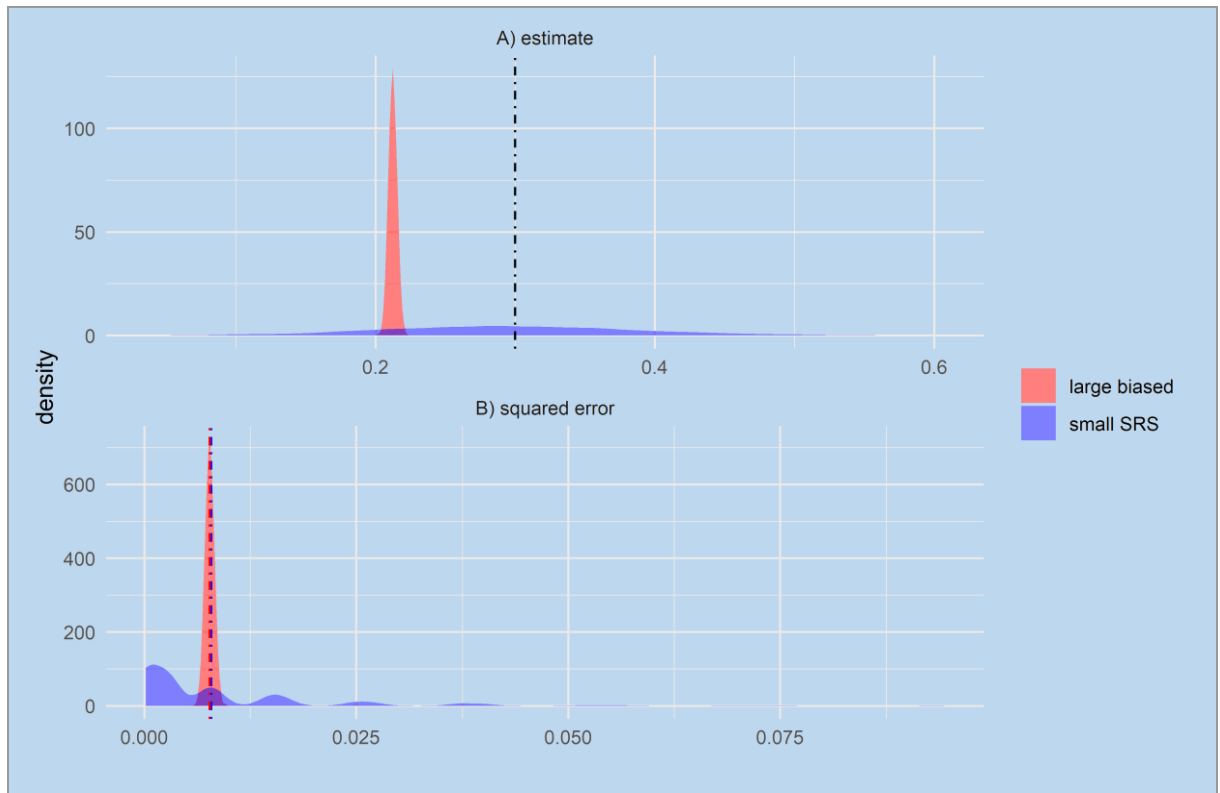

**Fig II.** Density plots showing the distributions of two statistics from 1000 simulated SRSs of size $n_{eff}$ = 28 (small SRS) and 1000 of size n = 19,419 with $\rho(R,Y)$ = -0.058 (large biased). Panel A shows the estimates of mean occupancy; the dashed line denotes the true value, 0.3. Panel B shows the squared differences between the sample-based estimates of mean occupancy and the population mean. It also shows the MSE from each set of simulations (dashed lines), which are very similar so hard to distinguish. These converge to the same value with increasing simulation number.

The interested reader should consult Meng [4], who demonstrated the big data paradox analytically. His analysis is based on the z score. Under the normal approximation, the z score denotes the half width of the confidence interval, in units of standard deviations, needed to obtain a given level of coverage. Meng shows that the z score needed to cover the true value is approximated by $\sqrt{n/n_{eff}}$, which is typically much larger than conventional values (e.g. 1.96 for 95% coverage). In our example, $\sqrt{n/n_{eff}}$ = 26.34.

All of the code and data needed to reproduce the analyses here and in Box 1 can be found at https://github.com/robboyd/selectionBiasEffects.

## Use and misuse of nonprobability samples in biodiversity monitoring

As we enter the era of "big data", ecologists have access to more (and larger) nonprobability samples than ever before. Examples include digitised museum and herbarium collections [11], distribution data collected for species atlases, citizen science data, and newer types of data from various sensors (e.g. acoustic and radar; [12]). Much of this data is available through data aggregators and meta-databases such as GBIF (gbif.org) and BioTIME [13]. Probability samples, too, may be held in data aggregators, but these become nonprobability samples when combined with additional data. Given the various challenges associated with inference from nonprobability samples, it is not surprising that the increasing availability of these for research has led to some high-profile disagreements in the biodiversity literature (Table 1).

Table 1. Nine recent examples of high profile biodiversity research papers with responses. We have chosen examples based on our reading rather than a systematic search for disagreements. Issues with which respondents voiced concerns have been divided by us into three main sampling domains: geography, the environment, and taxonomy/other [14]. The final column ("Issues mentioned...") categorises the original paper as to whether issues of sample representativeness were clearly acknowledged in either the title, abstract, or the paper's main body. The categories are: Y (Yes), N (No), and P (Partially). "Partial" recognition can either be due to the fact that unrepresentativeness was mentioned for some domains but not others, or because of a lack of recognition that a bias mitigation strategy was likely to have weaknesses, as highlighted by the given response. Note that we do not review any potential weaknesses in the "Responses" here, as we simply intend to demonstrate that there have been disagreements concerning sample coverage.

| Original paper | Response | Geographic domain | Environmental domain | Taxonomic/other domain | Issues mentioned in title/abstract/main body? |
|---|---|---|---|---|---|
| Bruelheide et al. [15] | Christensen et al. [16] | - | Change in the biotope (environmental domain) focus between the survey periods generated incomplete data. The method to handle these gaps while assessing trends was not sufficient, leading to a systematic underestimate of declines. | - | N/P/P |
| Crossley et al. [17] | Welti et al. [18] (W) Desquilbet et al. [19] (D) | - | Several datasets used to assess species trends over time are from experimental sites (for example, a site where a target species was being systematically removed) or have inconsistent sampling methods over time. | Both W and D, note that non-insect taxa were included in the analysis despite the taxonomic domain of interest being restricted to insects. | N/Y/Y |
| Hallmann et al. [20] | Saunders [21] (S) Vereecken et al. [22] (V) | S notes a lack of repeat sampling at the same locations potentially limits the ability to understand the true extent of declines. | S also notes that sites were restricted to small nature reserves. | V raise concerns about this paper (and Lister & Garcia [23]), where biomass is used as a proxy for declines in biodiversity and the services they provide. V show that the relationship between biomass and various biodiversity metrics/indicators | Y/P/Y |
| | | | | vary with habitat type. | |
| Leung et al. [24] | Murali et al. [25] | - | Murali et al. [25] note environmental bias in the populations included in the LPI, where populations inside protected areas are significantly over represented. Therefore the declines revealed in the LPI are likely to be worse at a global scale (the target domain of the LPI). | Murali et al. [25] also note that if the most extreme increasing vertebrate populations are excluded then the original LPI results remain broadly similar. | N/N/Y |
| WWF [26] | Leung et al. [24] | - | - | LPI estimates mean decline >50% for vertebrates since 1970. Leung et al. show that this is driven by <3% of the vertebrate populations included in the LPI; if excluded the mean trend becomes positive. | P* <br><br>*The LPI is a report: the P here refers to the main body of the report |
| Newbold et al. [27] | Martin et al. [28] | - | Concerns about apparent underestimated losses in the biodiversity intactness index, likely driven by significantly differing levels of human impact on the baseline "primary vegetation" sites used in the model to represent pristine condition. | - | N/N/P |
| Sánchez-Bayo and Wyckhuys [29] | Saunders [21] <br><br> Simmons et al. [30] | Title states "worldwide decline", replies highlight concerns around strong European and North American bias in the studies | - | Many insect groups are completely absent from the analysis, for example, cockroaches, termites and many fly and beetle families. Also the choice of search | N/N/P |

| | | included in the review. | | term (declin*) for the review could lead to bias in the subset of results, given the focus was "all long-term insect surveys conducted over the past 40 years…" | |
| --- | --- | --- | --- | --- | --- |
| Soroye et al. [31] | Guzman et al. [32] | Methodological issue where absence was inferred despite no evidence the location was visited, yielding biased estimates of decline. Furthermore, there was a large reduction in site visits between the two major time periods, particularly in North America and Southern Europe. | - | Species modelled across the entire geographic scope of the study, effectively assuming North American species may have been present at European sites, but had simply gone undetected, and vice versa for European bees at North American sites. | N/N/P |
| van Klink et al. [33] | Jähnig et al. [34] (J)<br><br>Audisio et al. [35] (A)<br><br>Scholl et al. [36] (S)<br><br>Murray-Stoker and Murray-Stoker [37] (M-S)<br><br>Desquilbet et al. [38] (D) | D highlights geographic bias in the modelled data with 76% of the studies covering the US and Europe, despite a global geographic domain of interest. J also note the restricted non-random geographic representation of the study. A total of five datasets cover Africa, South America and large parts of Asia. | S notes that many of the datasets included in the analysis are from studies examining insect responses to recent perturbation, and therefore are more likely to reflect population recovery. | J, A, S, M-S, all note that an increase in insect abundance should not be interpreted as a positive ecosystem response. This could be due to an increase in common, pollution tolerant species, while overall richness declines.<br><br>The domain of interest is insects ( the title stating "insect" abundance). However, D show crustacean, mollusc and worm data were included in the analysis. They also note that stress tolerant species were over-represented. | N/N/P |

The disagreements in Table 1 arguably all relate to issues of sample representativeness, with the possible exception of the reply to [31]. Soroye and colleagues [31] used an occupancy-detection model to estimate changes in the range sizes of bees in North America and Europe. Occupancy-detection models require data on detections and non-detections, and the published criticism related to how non-detections were inferred. We, too, are sceptical about the models of Soroye et al. [31], but suggest that the most pernicious issue is likely to be the lack of representativeness in their data, not the precise way that non-detections were inferred (cf. [32]). We should remember that any model-based estimates, whether of a data selection mechanism (e.g. an observation process) or of an ecological state variable, can suffer from bias. Occupancy-detection models use information from repeat visits to the same area to estimate the probability that a species is detected given that it is present, a form of measurement error. These estimates could themselves be biased if the sites or visit types are not representative of the typical data generating process [39]. We find that this is rarely, if ever, investigated by those applying occupancy-detection models to nonprobability samples, despite the fact that this is also a model of a real-world process.

The distinction between measurement error and representativeness can also be illustrated using Meng's formula for effective sample size (Box 1). The formula shows that, for a given selection bias, effective sample size is lower in the presence of measurement error [10]. This is not to say that removing measurement error would bring effective sample size back to parity with actual sample size. In our example (Box 1) we assume no measurement error, but a small amount of selection bias causes the effective sample size to be 99.86% lower than the actual sample size. This clearly demonstrates that dealing with measurement error alone does not fix issues caused by a lack of representativeness.

# How can we do better?

Given such high-profile disagreements, the potentially wasted research time [40,41], and the clear issues with small effective sample sizes and unrepresentativeness, one might reasonably argue that ecologists should stop using nonprobability samples for the purpose of monitoring biodiversity. Indeed, some researchers essentially argue from this standpoint, and it is not unusual to see such samples written off as "unscientific" (e.g. [42,43]). We do not fully accept this rather harsh view, although we have some sympathy with it. Elsewhere, statisticians working on survey sampling consider nonprobability samples as almost a fact of life rather than an indictment, and have embraced the resulting inferential challenges [5]. We suggest that many of the potential pitfalls of nonprobability samples could be avoided by formal assessments of the risk-of-bias, clear and honest communication, and mitigation (to the extent possible) of the unrepresentativeness that is typical of such samples in ecology. Much can be learned from other disciplines in each of these areas (see "Outstanding questions").

## Formal risk-of-bias assessments

The field of medical research has arguably led the way in terms of assessing and documenting the types of issues we have raised: qualitative risk-of-bias assessments, for example, are typically required for primary studies and evidence synthesis in this area (Table 2). These often focus on the impact of bias on causal inference (e.g. in studies on medical interventions; [44]), but the principle can equally be applied to sampling, where the potential for bias relates to systematic differences between sample and population, rather than between treatment and control [45]. Simons and colleagues [46], for example, proposed the inclusion of a "constraints on generality" statement within all primary psychological research papers, with the aim of highlighting the extent to which experimental findings on particular groups are likely to be generalizable. Boyd and colleagues [14] developed a similar initiative for biodiversity time trends, encouraging researchers to complete a formal "Risk-Of-Bias In Temporal Trends" (ROBITT) assessment when publishing such descriptive studies (see also [47] for a broader discussion of generality in ecology). Table 2 lists examples of similar risk-of-bias tools across disciplines, including information on uptake.

Table 2. Examples of qualitative risk-of-bias (RoB) tools for different types of primary study-level assessments across disciplines. V = version; citation numbers estimated by Google Scholar (1-Sep-2022) where not otherwise attributed.

| Tool | Field | Study/data type | Details | Community promotion | Reference(s) | Citations |
|---|---|---|---|---|---|---|
| Cochrane RoB tool | Medical research | Randomised controlled trials of medical interventions | Used to qualitatively stratify meta-analyses according to RoB | Assessment of RoB is regarded as an essential component of a systematic review in this area | V1: [48]; V2: [44] | >40,000 [44] |
| CEE Critical Appraisal Tool | Environmental management | Experimental/quasi-experimental studies | For qualitative assessment of RoB across different study designs used in the research area | To assist environmental evidence synthesisers conduct critical appraisal | V0.3 (prototype): [49] | 2 |
| Constraints On Generality (COG) tool | Psychology | Any inferential study | Engenders clear definition of the statistical population of interest and assesses external validity | Requested by some psychology journals as a part of good practice in open science [50] | [46] | 624 |
| PROBAST | Medical research | Predictive modelling studies of diagnoses and prognoses | Used at either the primary research or systematic review stage | Endorsed and recommended by journals in the area | [51][52] | 1136 |
| ROBINS-I | Medical research | Non-randomized studies of medical interventions | Compares data to that of a hypothetical randomised trial | Endorsed and recommended by journals in the area | [53] | 6787 |
| ROBITT | Ecology | Descriptive inference, especially temporal trends | Assessment of potential representativeness across relevant study domains | None to date | [14] | 2 |

## Mitigation

Statisticians working on survey sampling problems have developed methods to adjust for unrepresentativeness in nonprobability samples. These methods generally come under the banner of "**model-based inference**", which may be contrasted with the design-based approach often used for probability samples (model-based approaches can also be used with probability samples, and may be preferable to design-based inference in some cases; [54]). The key distinction between these modes of inference is the way in which the population is treated: in design-based inference, the population values are treated as fixed (the random element is the sampling); in model-based inference, the population values are treated as a realisation of a stochastic data-generating process [55,56]. If this data-generating process, the model, can be recovered from the sample, then it can be used to draw inferences about the population [6,57,58].

Of course, there is a risk that a model constructed from a sample will not extrapolate well to non-sampled population units. To increase the chances that the model is representative of non-sampled units, **design variables,** thought to explain the sample selection process, can be included as covariates [57]. Alternatively, models can be used to estimate sample inclusion probabilities; these can be used to construct design weights, enabling the researcher to proceed in a similar way to design-based inference [59]. This sometimes is referred to as "quasi-randomization" [6], and is similar to propensity score weighting in causal inference [45].

Some attempts have been made to correct for the unrepresentative nature of big biodiversity datasets. In the Living Planet Index—an indicator of vertebrate population trends—the contribution of each biogeographic region to the global trend is weighted by an estimate of regional species richness [60].

Johnston and colleagues [61] estimated the probability that birds were searched for across spatial units in Great Britain; they then used these estimates to weight the likelihood function when fitting a regression model predicting species' occupancy. Maxent, a popular species distribution modelling algorithm, includes an option to "factorBiasOut" [62]. The user provides information on sampling effort as input, and this is used to adjust the estimates of species' habitat suitability. These contributions are a good start for biodiversity science, but they need to become routine practice.

It should also be remembered that bias mitigation is not guaranteed to succeed, and that "success" can be very difficult to evaluate [5,6]. For example, Boyd and colleagues [63] found that SDMs fitted using a standard method to correct for sampling effort, the target group approach [64], were often rated as "poor" by taxon experts. In some circumstances, statistical correction procedures have even been shown to make SDM predictions worse [65].

## Communication

Even where bias mitigation strategies successfully reduce the risk-of-bias, it is unlikely that they will completely eliminate it: the remaining risk should be clearly communicated to readers and data users [66]. Issues of representativeness should also be conveyed in paper titles and abstracts, or at the least these should not be actively misleading through the omission of key qualifications (Table 1; "Outstanding questions"). An example of good practice would be to specify the geographic and taxonomic extents of a study in its title, rather than letting the reader assume that the conclusions are more widely applicable. Researchers who understand a dataset, and the phenomenon being studied, should be able to integrate both bias-related and ecological factors when discussing results (e.g. see [8]). The risk-of-bias could also be communicated graphically, as in health research [67]. Similar approaches have recently been proposed for use with ecological indicators and species trends [68].

Box 3. Glossary of terms.

> **Descriptive inference:** The process of estimating some quantity describing a population from a sample of that population. This can be broadly contrasted with other inferential goals, such as prediction or causation.
> **Design-based inference:** A mode of inference from sample to population. Design-based inference is most straightforward with probability samples, because sample inclusion probabilities can be used to adjust for selection bias. Design-based inference can be traced back to the work of Jerzy Neyman in the early- and mid-20$^{th}$ century.
> **Design variables:** Variables that influence, or are thought to influence, the probability that a population unit is sampled.
> **Design weights:** Often the inverse of sample inclusion probabilities. Design weights are used to correct for selection bias (as well as issues such as non-response) in design-based inference.
> **Estimator:** A rule for calculating an estimate of a population quantity from a sample. An oft-cited example is the sample mean, which is an estimator of the population mean.
> **Estimator bias:** A systematic deviation of a sample-based estimate from its corresponding population quantity. In a technical sense, this can imply different processes depending on the context, but it is always the irreducible, or "non-sampling", component of error; that is, it cannot be reduced simply by adding more of the same "type" of data.
> **Model-based inference:** A mode of inference from sample to population. In model-based inference, a model thought to describe the variable of interest in the population is constructed, and inferences are drawn from this model. Model-based inference can be traced back to the work of Ronald Fisher in the early 20$^{th}$ century.
> **Representativeness:** A common definition of sample representativeness is the correlation between an indicator variable, taking the value 1 if a population unit is in the sample and 0 otherwise, and the outcome variable of interest. Larger correlations equal lower representativeness. A representative sample has little selection bias, and vice versa. In some areas, representativeness is assessed using "R-indicators", which are measures of the variability of sample inclusion probabilities.

> **Selection bias:** Selection bias induces a correlation between an indicator variable, taking the value 1 if a population unit is in the sample and 0 otherwise, and the variable of interest. Hence, selection bias results in unrepresentative samples. If not dealt with, selection bias leads to **estimator bias**.

## Concluding remarks

The word science ultimately derives from the Latin verb *scire*, "to know", but do the types of biodiversity studies and approaches discussed here really yield accurate knowledge? It is difficult to know unless, as readers, we are presented with honest appraisals of the potential risk-of-bias and appropriate sensitivity analyses. As the statistician John Tukey opined towards the end of his career, *"[t]he combination of some data and an aching desire for an answer does not ensure that a reasonable answer can be extracted from a given body of data"* [69]. Papers, their titles and their abstracts should all represent this uncertainty, and journals should not give authors a free pass on these in the pursuit of "impact" and publicity. This will likely require a step change in the incentive structure for scientific publishing and funding ([40]; see "Outstanding questions").

## Outstanding questions

**Is the use of nonprobability samples cost-effective?** It could be argued that we are in a vicious cycle: if researchers continue to assert that it is possible to monitor biodiversity accurately using highly unrepresentative samples, then probability sampling is less likely to be funded.

**Will new types of biodiversity data—e.g. from acoustic, radar and other sensors—make issues of representativeness better or worse?** These sensors have the potential to produce unprecedented quantities of data, but are likely to suffer from issues of representativeness. For example, acoustic monitoring stations could still be placed in an unrepresentative set of locations, and radar can only detect species above a particular size.

**How can we incentivise communication of the risk-of-bias and unrepresentativeness in scientific papers?** High impact journals seek eye-catching results, sometimes at the expense of rigour and clarity. If understanding the true state of our knowledge about reality is the goal, then the communication of such uncertainties has to improve.

**What other methods for making inferences from nonprobability samples exist, and how reliable are they with real data?** Disciplines such as political science, epidemiology and applied statistics have investigated methods for making inferences from nonprobability samples (or analogous methods for causal estimands). How many are likely to be suitable for ecology, and what should users report to fully communicate their likely final accuracy to the reader?

**What other opportunities does the honest representation of uncertainty in biodiversity science present?** For example, expert assessments of the representativeness of aggregated biodiversity data, or other nonprobability samples, may provide additional opportunities for taxon and field-based experts to contribute to the ongoing data science revolution.

## Acknowledgements

Thank you to Xiao-Li Meng, Andrea Stephens, Robin Walls, and three anonymous reviewers whose comments all improved the paper. We also thank Robin Hutchinson for facilitating access to the *Calluna* iRecord data, and the Botanical Society of Britain and Ireland for providing access to their *Calluna* Atlas data. All authors were supported by the NERC award number NE/R016429/1 as part of the UK Status, Change and Projections of the Environment (UK-SCAPE) program delivering National Capability.